\documentclass[structabstract]{aa}  

\usepackage{graphicx}
\usepackage{txfonts}
\usepackage{natbib}
\usepackage{longtable}
\usepackage{mathptmx}

\begin{document}

\title{Astrometric suitability of optically-bright ICRF sources \\
for the alignment with the future Gaia celestial reference frame}

\author{G. Bourda
\and P. Charlot
\and J.-F. Le Campion}

%\titlerunning{Astrometric suitability of ICRF sources for the alignment with Gaia}
\titlerunning{Draft paper to appear in Astronomy \& Astrophysics}
\authorrunning{G. Bourda et al.}

\offprints{G. Bourda}

\institute{Laboratoire d'Astrophysique de Bordeaux - Universit\'{e} de Bordeaux - CNRS UMR5804, BP89, 33271 Floirac Cedex, France \\
e-mail: Geraldine.Bourda@obs.u-bordeaux1.fr}

\date{Received July 24, 2008; accepted August 1, 2008}

\abstract {The International Celestial Reference Frame (ICRF), currently based on the position of 717 extragalactic radio sources observed by VLBI (Very Long Baseline Interferometry), is the fundamental celestial reference frame adopted by the International Astronomical Union (IAU) in 1997. Within the next 10 years, the European space astrometry mission Gaia, to be launched by 2011, will permit determination of the extragalactic reference frame directly in the visible for the first time. Aligning these two frames with the highest accuracy will therefore be very important in the future for ensuring consistency between the measured radio and optical positions.} {This paper is aimed at evaluating the current astrometric suitability of the individual ICRF radio sources which are considered appropriate for the alignment with the future Gaia frame.} {To this purpose, we cross-identified the ICRF and the optical catalog V\'{e}ron-Cetty and V\'{e}ron (2006), in order to identify the optically-bright ICRF sources that will be positioned with the highest accuracy with Gaia. Then we investigated the astrometric suitability of these sources by examining their VLBI brightness distribution.} {We identified 243 candidate ICRF sources for the alignment with the Gaia frame (i.e. with an optical counterpart brighter than the apparent magnitude 18), but only 70 of these (i.e. only 10\% of the ICRF sources) are found to have the necessary high astrometric quality (i.e. a brightness distribution that is compact enough) for this link. Additionally, it was found that the QSOs (quasi stellar objects) that will have the most accurate positions in the Gaia frame tend to have less-accurate VLBI positions, most probably because of their physical structures.} {Altogether, this indicates that identifying other high-quality VLBI radio sources suitable for the alignment with the future Gaia frame is mandatory.}


\maketitle

%%%%%%%%%%%%%%%%%%%%%%%
\section{Introduction}

The International Celestial Reference System (ICRS) is a kinematical system that assumes that the Universe does not rotate, hence breaking the history of the inertial system materialization from observations attached to the apparent motion of the Sun \citep{Arias1995}. The International Celestial Reference Frame (ICRF) is the realization at radio wavelengths of the ICRS, through very long baseline interferometry (VLBI) measurements of extragalactic radio source positions. It was adopted by the International Astronomical Union (IAU) as the fundamental celestial reference frame during the IAU XXIIIrd General Assembly in Kyoto, Japan, in August 1997. The initial realization was based on the positions measured with VLBI of 212 \textit{defining} extragalactic radio sources (setting the direction of the ICRF axes). In addition, positions for 396 other sources (divided into 294 \textit{candidate} and 102 \textit{other} sources) were included to make the frame denser \citep{Ma1998}. The ICRF was extended at later stages with the positions of another 109 sources commonly referred to as \textit{new} sources \citep{Fey2004}. Overall, the ICRF currently consists of a set of VLBI coordinates for 717 extragalactic radio sources, most of which show sub-milliarcsecond accuracy. The accuracy of the individual source positions depends on the number of observations but also on the compactness and positional stability of the sources that show brightness distributions that are often not point-like and time-variable (\citealt{Ma1998}; \citealt{Fey2000}).

The European space astrometry mission Gaia, to be launched by 2011, will survey about one billion stars in our Galaxy and throughout the Local Group along with 500\,000 quasi stellar objects (QSOs), down to an apparent optical magnitude of 20 \citep{Perryman2001}. Optical positions with Gaia will be determined with an unprecedented accuracy, ranging from a few tens of microarcseconds ($\mu$as) at magnitude 15--18 (targeted accuracies are 16 $\mu$as at 15 mag and 70 $\mu$as at 18 mag) to about 200~$\mu$as at magnitude 20 \citep{Lindegren2008}. Unlike Hipparcos, Gaia will permit the construction of an extragalactic frame directly at optical wavelengths, based on the QSOs with the most accurate positions (i.e. the QSOs with magnitude brighter than 18; \citealt{Mignard2003}). \citet{Mignard2002} demonstrates that the residual spin of the Gaia frame can be determined to $0.5~\mu$as/yr with a ``clean sample'' of about 10\,000 such QSOs. A preliminary Gaia catalog is expected to be available by 2015 with the final version released by 2019.

In the future, aligning the ICRF with the Gaia celestial reference frame will be crucial for ensuring consistency between the measured radio and optical positions. This alignment will be important not only for guaranteeing the proper transition in the case that the fundamental reference frame is moved from the radio domain to the optical domain (as currently anticipated) but also for registering the radio and optical images of any celestial target with the highest accuracy. To align the two frames with the highest accuracy, it is desirable to have several hundred common objects (the more common objects, the more accurate the alignment) with a uniform sky coverage and very accurate radio and optical positions. Obtaining such accurate positions implies that the link sources must have (i) an apparent optical magnitude $V$ brighter than 18, for the highest Gaia astrometric accuracy \citep{Mignard2003}, and (ii) no extended VLBI structures, for the highest VLBI astrometric accuracy \citep{Fey2000}.

This paper is aimed at evaluating the suitability of the current individual ICRF extragalactic radio sources for the alignment with the future Gaia frame. First, the ICRF is cross-identified with the \citet{Veron2006} optical catalog of QSOs in order to identify the ICRF sources with a proper optical counterpart for the Gaia link. The astrometric suitability of these sources is then investigated by examining their VLBI brightness distribution and their position accuracy in the ICRF. In this investigation, the structure index defined by \citet{Fey2000} is used to identify the sources that have the most compact VLBI brightness distributions. This study results in a sample of 70 ICRF sources, which have appropriate compact structures on VLBI scales and are brighter than magnitude 18. These sources are at present the best candidates for the ICRF--Gaia alignment.

%%%%%%%%%%%%%%%%%%%%%%%%%%%%%%%%%%%%%%%%%%%%%%
\section{Optical counterparts of ICRF sources}

%---------------------------------------------------------------
\subsection{Cross-identification of radio and optical positions}

\citet{Veron2006} compiled several optical surveys of active galactic nuclei (AGN) from the literature in order to create a homogeneous, compact, and convenient catalog of 108\,080 objects (85\,221 quasars, 1\,122 BL Lac objects, and 21\,737 active galaxies). To identify the ICRF sources that have an optical counterpart, we have cross-identified the positions in this catalog with those in the ICRF. For this cross-identification, it was necessary to adopt an upper limit on the optical-radio position differences above which the corresponding radio and optical sources were considered as different objects. 

We tested the impact of the selected upper limit by carrying out the cross-identification of the two catalogs for several values of this parameter, ranging from $1^{\prime\prime}$ to $5^{\prime\prime}$, the results of which are shown in Table~\ref{table:tab1}. From this table, it is found that there are no major differences whether one sets the upper limit value to $3^{\prime\prime}$, $4^{\prime\prime}$ or $5^{\prime\prime}$. On the other hand, significantly fewer sources are found to be common when an upper limit value of $1^{\prime\prime}$ or $2^{\prime\prime}$ is selected. Finally we decided to adopt the upper limit value of $3^{\prime\prime}$, equivalent to three times the accuracy of the optical positions in the \citet{Veron2006} catalog, which is roughly at the $1^{\prime\prime}$ level. Uncertainties in the VLBI positions make a negligible contribution to the position differences since such uncertainties are at the milliarcsecond level. 

Based on this cross-identification, we determined that 75~\% of the ICRF sources (i.e. 535 sources out of a total of 717 ICRF sources) have an optical counterpart in \citet{Veron2006}.This set of sources includes 406 quasars (76~\%), 65 BL Lac objects (12~\%), and 64 active galaxies (12~\%). These results are comparable to those of \citet{Souchay2006} who used the \citet{Veron2003} optical catalog for the cross-identification.
 
Table~\ref{table:tab2} presents the distribution of the 535 cross-identified ICRF sources, depending on their categorization in the ICRF as \textit{defining}, \textit{candidate}, \textit{other} or \textit{new} sources. This table shows that:
\begin{itemize}
\item  83~\% of the ICRF \textit{defining} sources, 70~\% of the ICRF \textit{candidate} sources, 88~\% of the ICRF \textit{other} sources, and 60~\% of the ICRF \textit{new} sources have an optical counterpart in \citet{Veron2006}.  
\item  33~\% of the ICRF sources having an optical counterpart in \citet{Veron2006} are \textit{defining} sources, 38~\% \textit{candidate} sources, 17~\% \textit{other} sources, and 12~\% \textit{new} sources.
\end{itemize}
Based on these statistics, there is no evidence for any dependence of the existence of optical counterparts on ICRF source categories. The sample of sources identified here forms the initial pool of ICRF sources from which to identify the best candidates for the Gaia link after consideration of optical magnitudes, source structure, and astrometric accuracy. 

\begin{table}
\caption{Number of ICRF extragalactic radio sources cross-identified with the optical catalog of V\'{e}ron-Cetty and V\'{e}ron (2006; noted VV2006), depending on the source type in this catalog and on the upper limit used for the position differences in the cross-identification.}
\label{table:tab1}     
\centering                         
\begin{tabular}{llrrrrr}        
\hline\hline 
\noalign{\smallskip}               
   &  & $5^{\prime\prime}$ & $4^{\prime\prime}$ & $3^{\prime\prime}$ & $2^{\prime\prime}$ &  $1^{\prime\prime}$ \\
\noalign{\smallskip}               
\hline                       
\noalign{\smallskip}               
ICRF & $\cap$ Quasars                  & 409 & 409 & 406 & 387 & 250 \\
\noalign{\smallskip}               
     & $\cap$ BL Lac                   &  66 &  65 &  65 &  64 &  45 \\
\noalign{\smallskip}               
     & $\cap$ Active galaxies          &  66 &  66 &  64 &  61 &  36 \\
\noalign{\smallskip}               
\hline                       
\noalign{\smallskip}               
\multicolumn{2}{l}{ICRF $\cap$ VV2006} & 541 & 540 & 535 & 512 & 331 \\
\noalign{\smallskip}               
\hline                       
\end{tabular}
\end{table}

\begin{table}
\caption{Number of ICRF extragalactic radio sources cross-identified with the optical catalog of V\'{e}ron-Cetty and V\'{e}ron (2006; noted VV2006), depending on the ICRF source category.}
\label{table:tab2}     
\centering                          
\begin{tabular}{lrrrrr}        
\hline\hline 
\noalign{\smallskip}               
                          &\multicolumn{4}{c}{ICRF catalog} \\
\noalign{\smallskip}               
                          &\textit{Defining}&\textit{Candidate}&\textit{Other}&\textit{New}\\
                          &     (212) &     (294) & (102) & (109) \\
\noalign{\smallskip}               
\hline                       
\noalign{\smallskip}               
VV2006 $\cap$ ICRF        &           &           &       &       \\
\noalign{\smallskip}               
~~~~Quasars (406)         &    132    &   160     &  70   &  44   \\
\noalign{\smallskip}               
~~~~BL Lac (65)           &     27    &    24     &  11   &   3   \\
\noalign{\smallskip}               
~~~~Active galaxies (64)  &     16    &    21     &   9   &  18   \\
\noalign{\smallskip}               
\hline                       
\noalign{\smallskip}               
Total (535/717)           &   175/212 &  205/294  & 90/102& 65/109\\
\noalign{\smallskip}               
\hline  
\end{tabular}
\note{The upper limit used for the position differences in the cross-identification is $3^{\prime\prime}$.}
\end{table}

%----------------------------------------------
\subsection{Distribution of optical magnitudes}

\begin{figure}
\centering
\includegraphics[angle=-90,width=9cm]{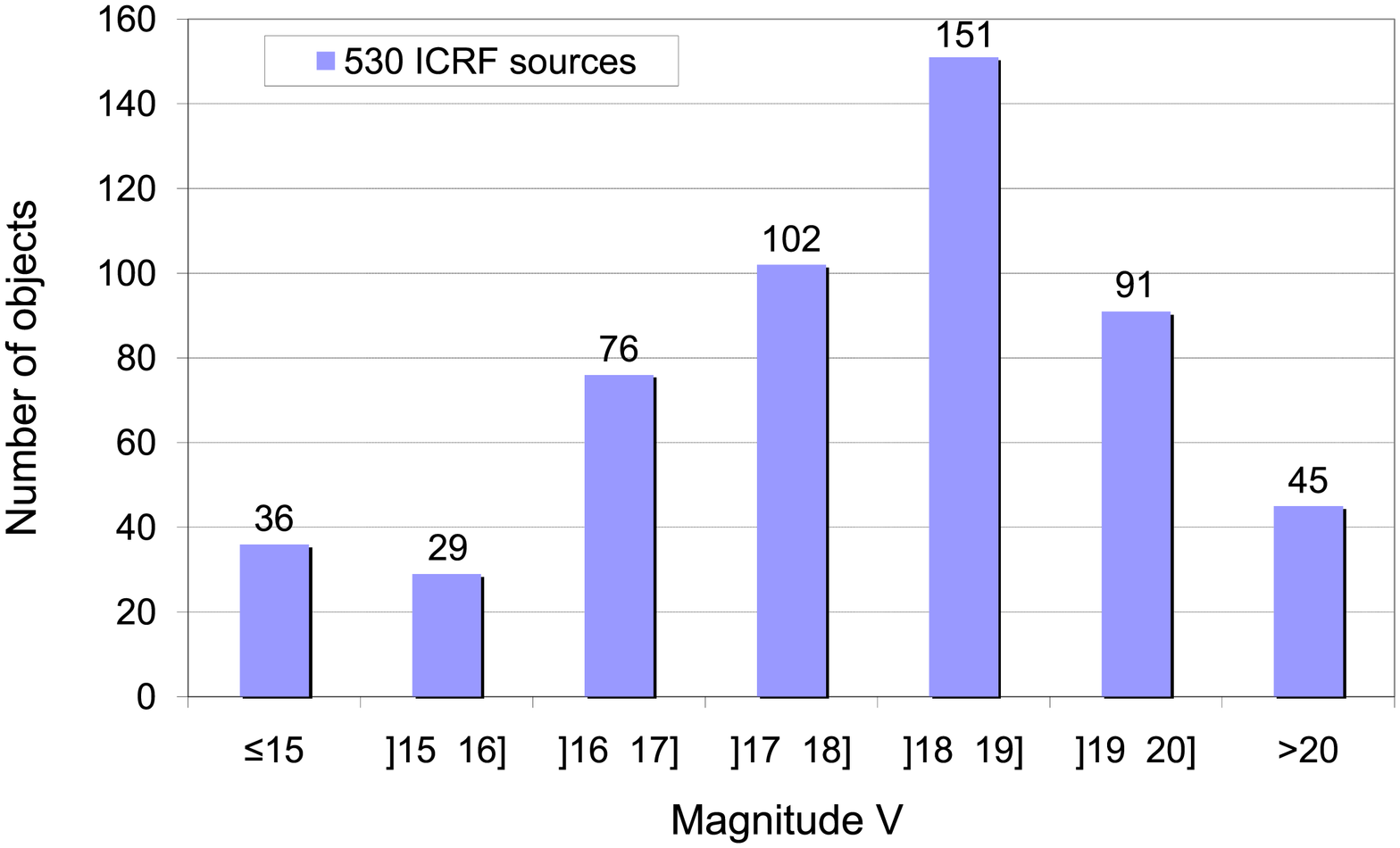}
\caption{Distribution of the apparent optical magnitude $V$ for the 530 ICRF extragalactic radio sources that have a \citet{Veron2006} optical counterpart.}
\label{Figure:Fig1}
\end{figure}

In a second stage, we investigated the distribution of the apparent optical magnitude $V$ for the 535 ICRF sources identified in the previous step. Five of these sources, with undefined optical magnitudes in the \citet{Veron2006} catalog, have not been considered. As shown in Fig.~\ref{Figure:Fig1}, the distribution of optical magnitudes peaks at $V$ between 18 and 19 with the bulk of the sources brighter than $V=20$. Overall, there are 485 sources brighter than 20, i.e. detectable by Gaia, while only 45 sources are weaker than 20. When restricting the magnitude limit to 18, the sample is reduced by a factor of 2, leaving only 243 sources. These 243 sources, representing 34~\% of the ICRF catalog, would be good candidates for establishing the alignment with the future Gaia celestial reference frame, provided they show limited structure on VLBI scales and have high VLBI positional accuracy.

%%%%%%%%%%%%%%%%%%%%%%%%%%%%%%%%%%%%%%%%%%%%%%%%%%%%%%%%%%%%%%%%%%
\section{Astrometric suitability based on observed VLBI structure}

The quality of the alignment between the ICRF and the Gaia frame will depend strongly on the astrometric suitability of the sources used for this alignment. To ensure the highest accuracy, the sources with the highest astrometric quality from the above sample should be used. 

On the milliarcsecond scale, most of the extragalactic radio sources exhibit spatially extended structures that vary in both time and frequency. As shown by \citet{Charlot1990}, such structures may introduce significant errors in the VLBI measurements that deteriorate the source position accuracy. The astrometric suitability of the sources may be estimated from the observed structures through the so-called ``structure index'', first introduced by \citet{Fey1997}. The structure index ranges from 1 to 4. Structure index values of 1 or 2 point to the most compact sources with the highest astrometric quality, whereas higher values of the structure index (3 or 4) correspond to sources with extended structures, which are less suitable for astrometry. For the alignment of the ICRF and Gaia frame, only sources with structure index values of 1 or 2, i.e. with the highest astrometric quality, should be used since one wants to determine the link with the highest accuracy.

\begin{figure*}[htbp]
\centering
\begin{minipage}[b]{1.\linewidth}
\centering
	\begin{minipage}[b]{.49\linewidth}	
	\centering
	\includegraphics[scale=0.35]{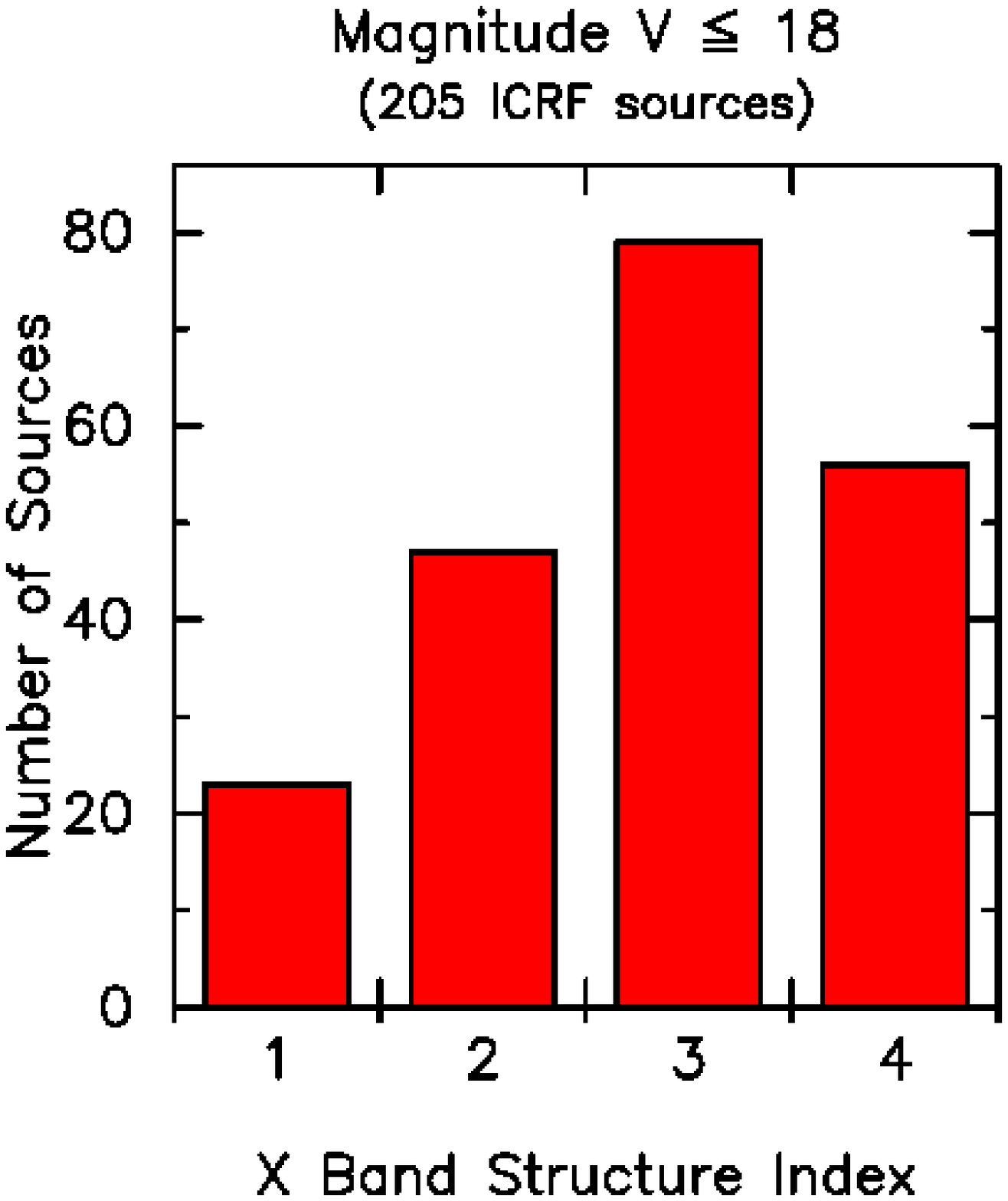}
	\vfill
	\includegraphics[scale=0.35]{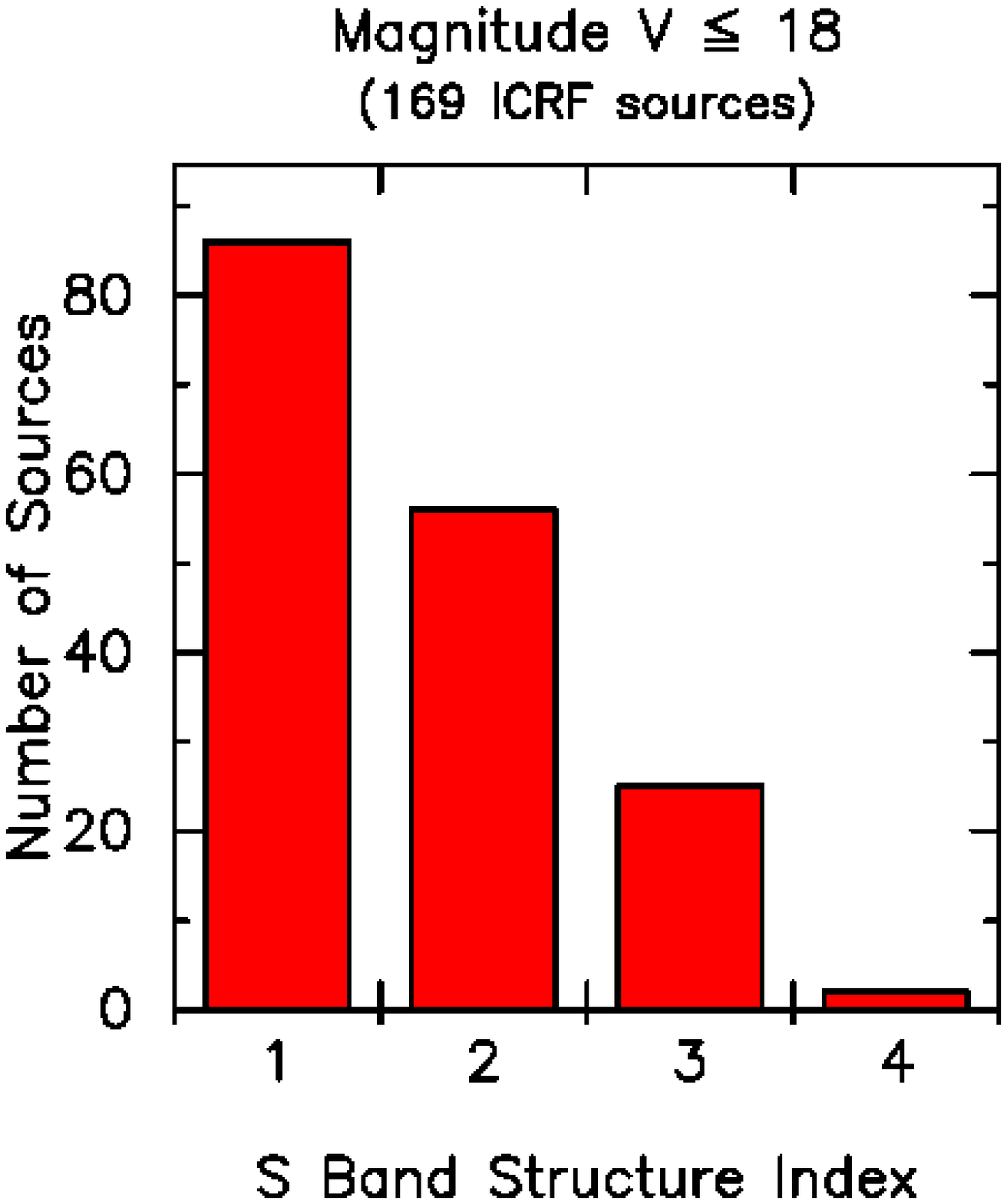}
	\end{minipage}
\hfill
	\begin{minipage}[b]{.49\linewidth}
	\centering
	\includegraphics[scale=0.35]{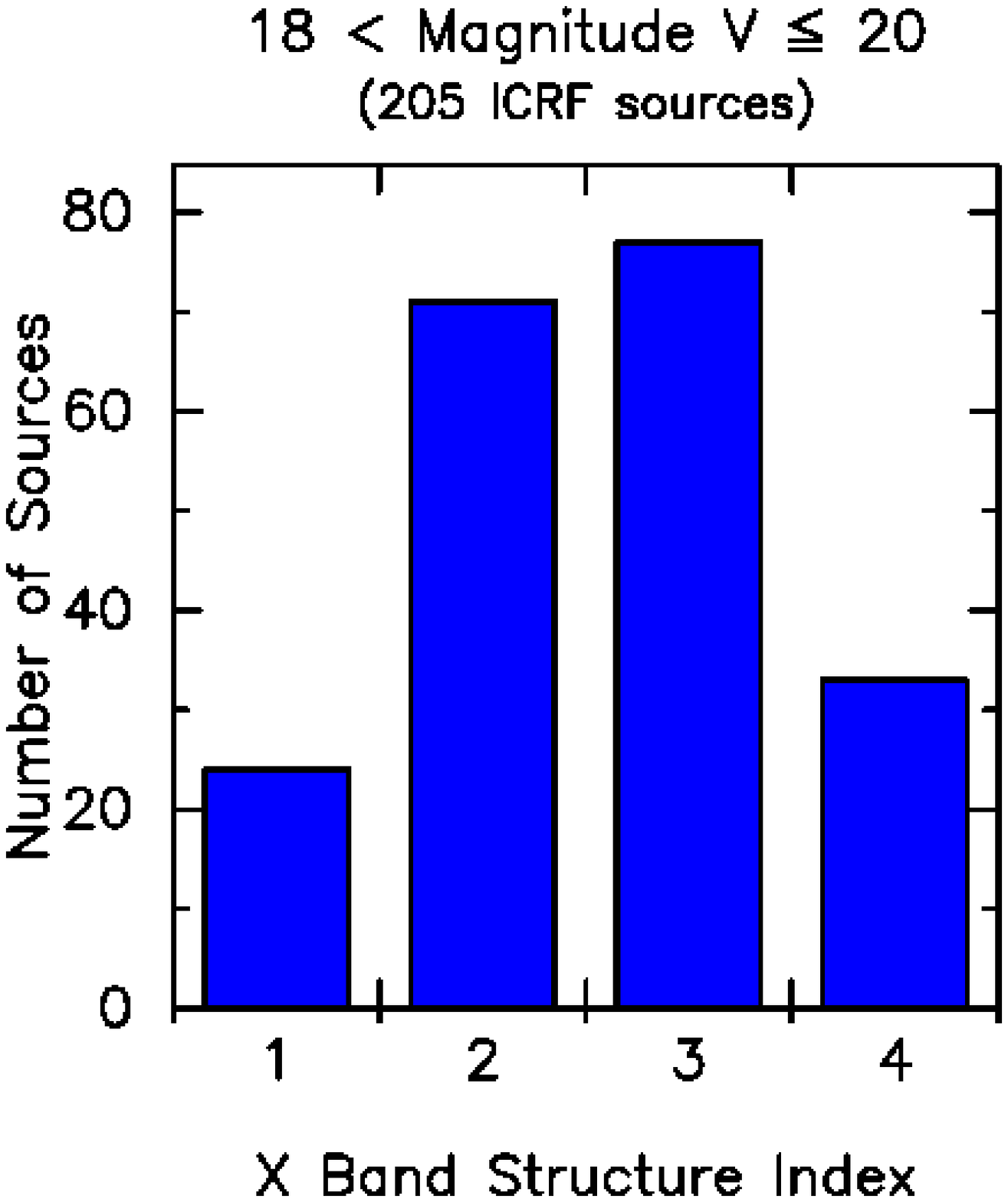}
	\vfill
	\includegraphics[scale=0.35]{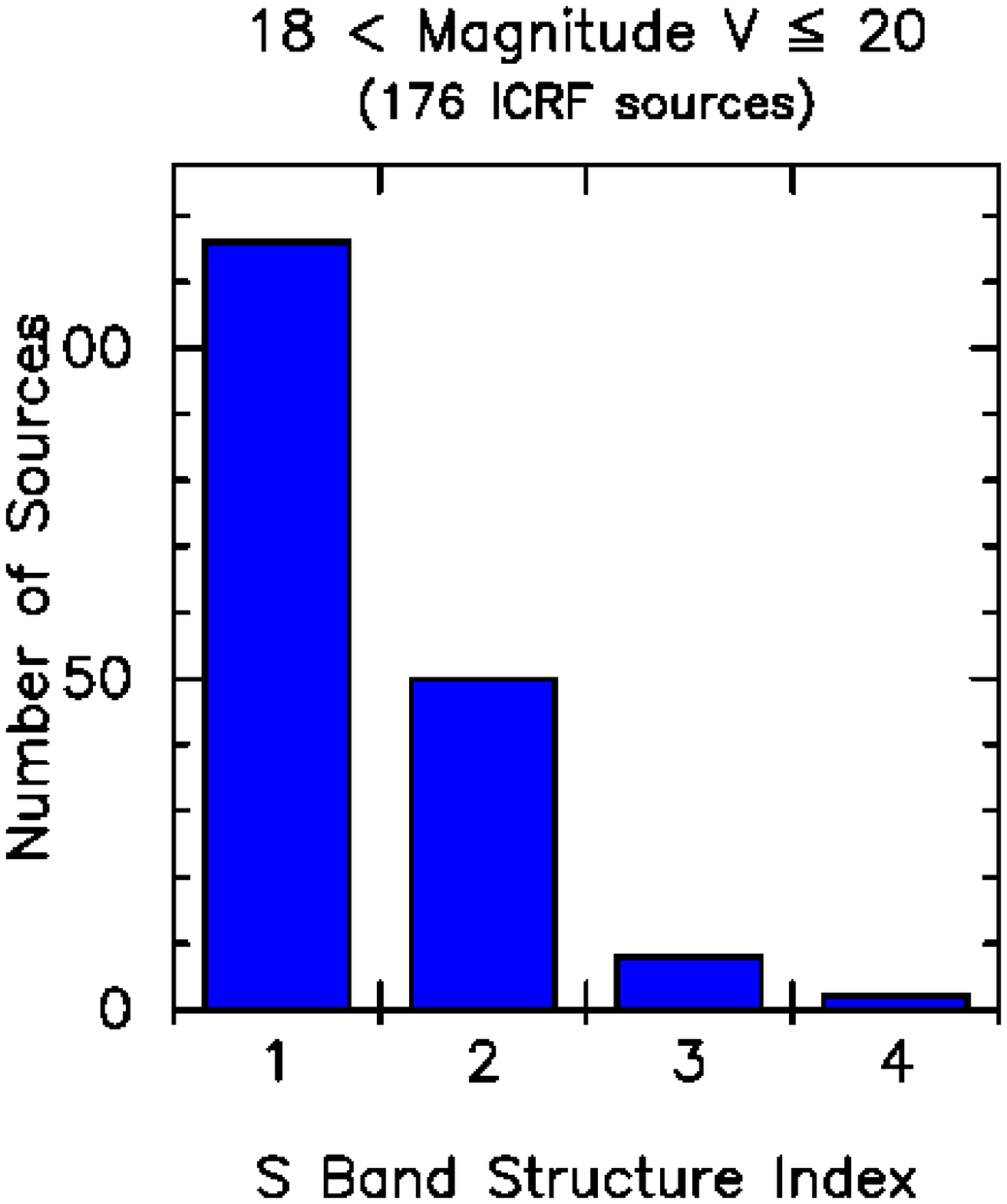}
	\end{minipage}
\end{minipage}
\caption{Distribution of the X- and S-band structure indices among the optically bright ICRF sources; upper panel: X-band structure index distribution; lower panel: S-band structure index distribution; left-hand side: sources with $V \leq 18$; right-hand side: sources with $18 < V \leq 20$.}
\label{Fig:Fig2}
\end{figure*}

\begin{figure}
\centering
\includegraphics[scale=0.4,angle=-90]{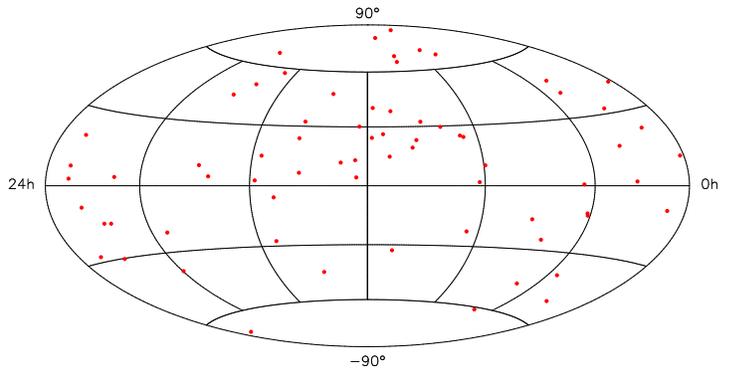}
\caption{Sky distribution for the 70 ICRF sources suitable for the alignment with the future Gaia frame.}
\label{Fig:Fig3}
\end{figure}

The structure index distribution for the 485 optically-bright ICRF sources with an apparent optical magnitude $V \leq 20$ (see Sect.~2) was investigated and is plotted separately for the sources with $V \leq 18$ and for those with $V$ between 18 and 20 in Fig.~\ref{Fig:Fig2}. For these plots, structure indices from the most recent calculation by \citet{Charlot2006} were used, and the distribution is shown for both S and X bands, the two standard frequency bands used in VLBI astrometry. Overall, structure index values are available for 410 sources at X band (205 sources with $V \leq 18$ and 205 sources with $18 < V \leq 20$) and 345 sources at S band (169 sources with $V \leq 18$ and 176 sources with $18 < V \leq 20$). Interestingly, the sources with magnitude $V \leq 18$ tend to have larger structure indices than those with $18 < V \leq 20$, as noticed when comparing the corresponding distributions in Fig.~\ref{Fig:Fig2}. This indicates that the brighter optical sources have statistically more extended radio structures. Based on this analysis, it is also found that only 70 ICRF sources out of those with $V \leq 18$ have an excellent or good astrometric suitability (i.e. an X-band structure index value of either 1 or 2) appropriate for determining the Gaia link with the highest accuracy (see list in Table~\ref{table:tab3}). This represents only a small fraction (about 10~\%) of the entire ICRF catalog. The sky distribution of these 70 sources, plotted in Fig.~\ref{Fig:Fig3}, is reasonable but indicates that there is a general lack of sources. 

\addtocounter{table}{1}

%%%%%%%%%%%%%%%%%%%%%
\section{Discussion}

\begin{figure}
\centering
\includegraphics[scale=0.6]{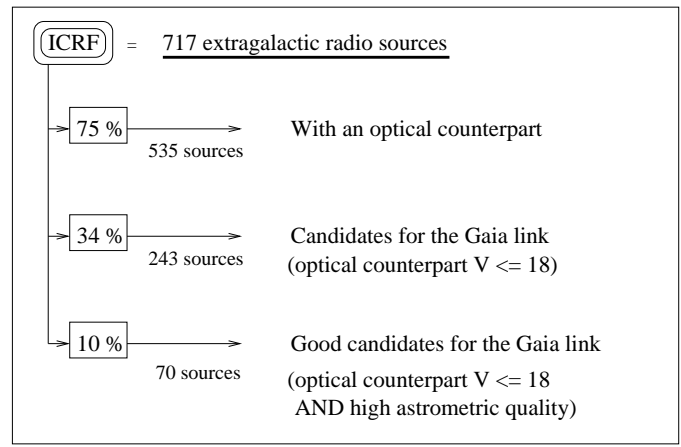}
\caption{Summary of the successive selection steps conducted to identify the best ICRF sources candidates for the alignment between the current ICRF and the future Gaia frame.}
\label{Fig:Fig4}
\end{figure}

This paper is aimed at identifying from the current ICRF extragalactic radio sources the best candidates to establish an accurate alignment with the future Gaia frame. Figure~\ref{Fig:Fig4} summarizes the successive steps completed to achieve this goal. We first pointed out that 75~\% of the ICRF sources have an optical counterpart (535 sources), while less than half of these are potential candidates for the alignment with the future Gaia frame (243 ICRF sources with $V \leq 18$, i.e. with the best positions measured by Gaia). And finally, we concluded that only 10~\% of the ICRF sources can be identified currently as the best candidates to establish this alignment with the highest accuracy (70 optically-bright ICRF sources with the highest astrometric quality). 

\begin{figure}
\begin{minipage}[b]{1.\linewidth}
\centering
\includegraphics[height=8cm,width=7.5cm]{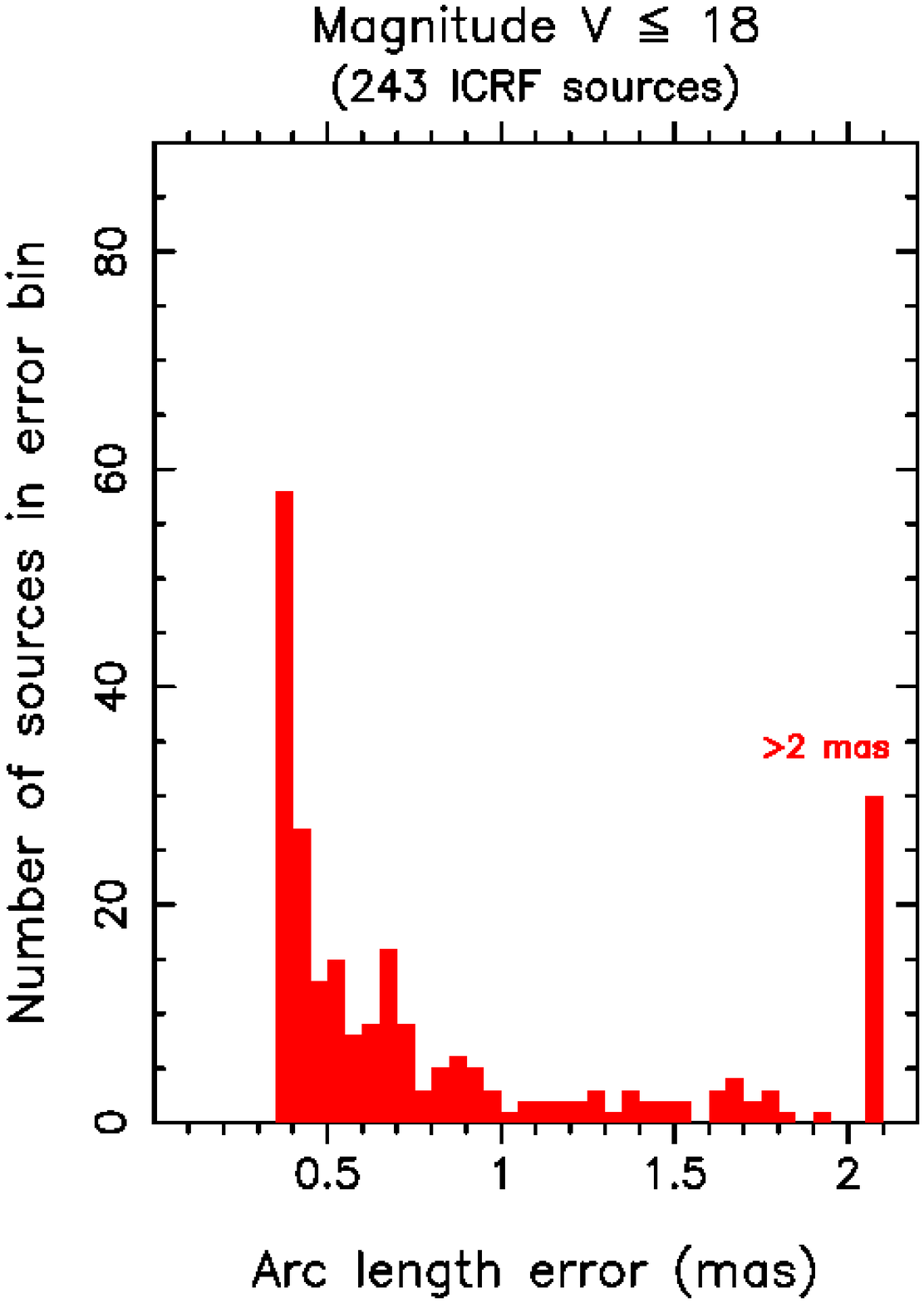}
\vspace{0.55cm}
\vfill
\includegraphics[height=8cm,width=7.5cm]{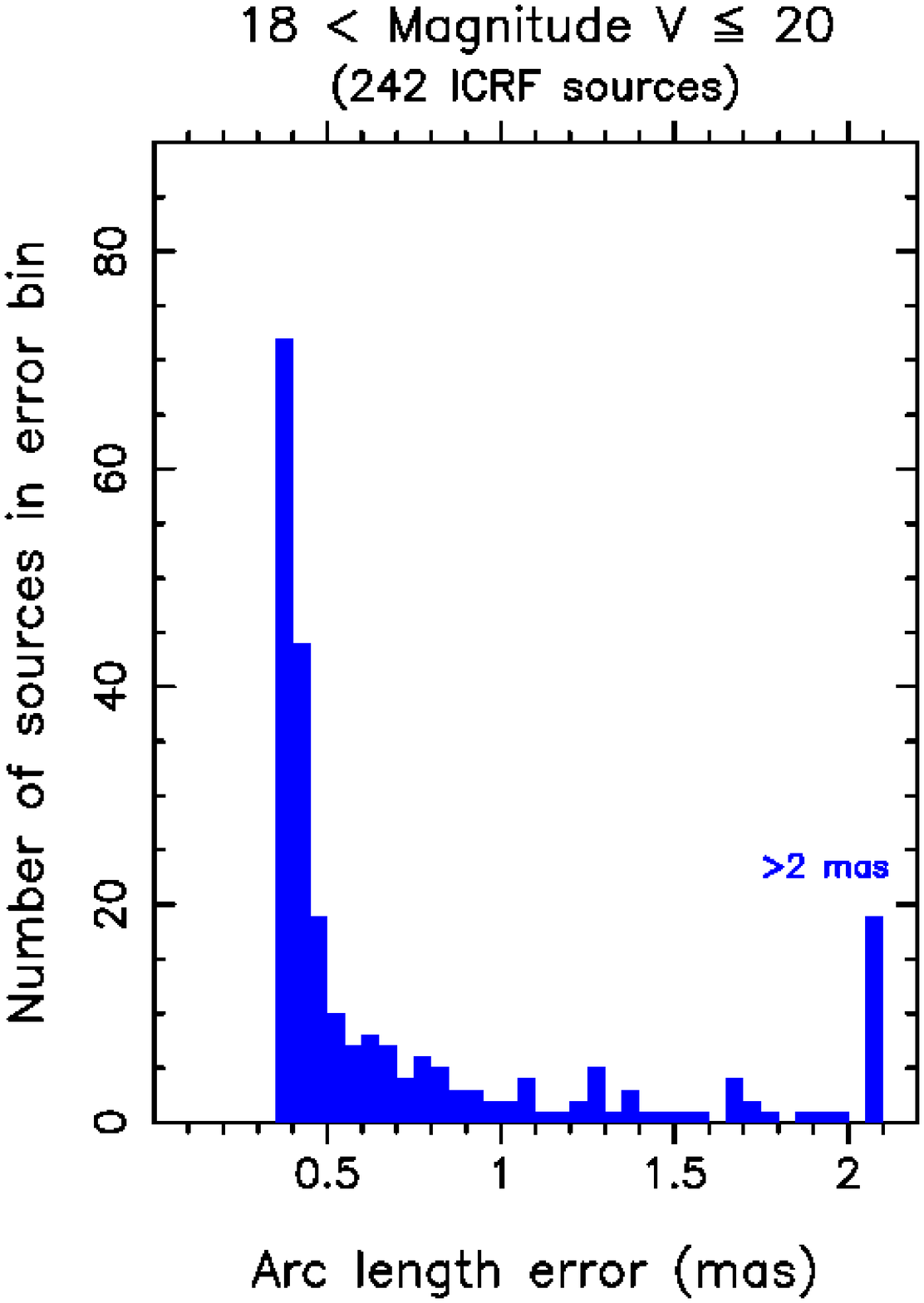}
\end{minipage}
\caption{Distribution of position accuracy for the 485 ICRF sources with an optical apparent magnitude $V$ brighter than 20. The accuracy is calculated as $\sqrt{\sigma^2_{\alpha}~cos^2 \delta + \sigma^2_{\delta}}$ where $\sigma_{\alpha}$ and $\sigma_{\delta}$ are the errors in right ascension and declination, as taken from the ICRF catalog. The upper panel shows the positional accuracy for the 243 sources with $V \leq 18$, while the lower panel shows the positional accuracy for the 242 sources with $18 < V \leq 20$.}
\label{Fig:Fig5}
\end{figure}

To supplement this study, we also investigated the distribution of position accuracy, as available from \citet{Ma1998} and \citet{Fey2004}, which is plotted in Fig.~\ref{Fig:Fig5} for the 485 optically-bright ICRF radio sources with $V \leq 20$. From this figure, we can notice that the sources with $18 < V \leq 20$ appear to have higher positional accuracy than those with $V \leq 18$. This is especially noticeable when comparing the fraction of sources with position accuracy better than 0.5 mas and those with positional accuracy in the range 0.5--1 mas. This fraction is 71\% for the sources with $18 < V \leq 20$ (135~sources out of a total of 190~sources with position accuracy better than 1~mas), whereas it is 55~\% for the sources with $V \leq 18$ (98 sources out of a total of 177 sources with position accuracy better than 1~mas). This trend is also confirmed when comparing the median positional accuracy, which is 0.60~mas for the sources with $V \leq 18$ and 0.47~mas for the sources with $18 < V \leq 20$. Such a finding is consistent with our results on astrometric suitability (Sect.~3), which indicate that the brighter sources ($V \leq 18$) have statistically more extended VLBI structures. It is thus likely that the deteriorated position accuracy for the $V \leq 18$ sources in Fig.~\ref{Fig:Fig5} is a result of their having larger structures. A similar correlation between source structure and position accuracy was also reported by \citet{Fey2000}. In the context of Gaia, this means that the sources that will have the most accurate optical positions measured by Gaia are not statistically those that have the best measured positions in the ICRF. It will thus be crucial in the future to unveil more high-quality optically-bright radio sources suitable for the Gaia link beyond those currently identified from the ICRF and reported in Table~\ref{table:tab3}.

A first opportunity to find such sources may come from the second generation of the ICRF (ICRF-2), currently under development, which will improve the position accuracy of the current ICRF sources and provide VLBI positions of additional sources. This new frame will rely on data acquired by international observing programs organized through the International VLBI Service for Geodesy and Astrometry \citep{Schluter2007} or other programs such as that developed in the southern hemisphere by the United States Naval Observatory \citep{Ojha2004,Ojha2005}. 
Another option would be to use the Very Long Baseline Array (VLBA) Calibrator Survey (VCS), which comprises milliarcsecond accurate positions for about 3000 additional sources along with images for a large fraction of those sources \citep{Beasley2002,Fomalont2003,Petrov2005,Petrov2006,Kovalev2007,Petrov2008}.
Finally, an additional possibility is to search for new VLBI sources, by targeting radio sources weaker than those observed so far in VLBI astrometry (i.e. with flux densities typically below 100~mJy). Such sensitive observations can now be realized owing to recent increases in the VLBI network sensitivity (e.g. recording at 1~Gb/s) and by using a network comprising large antennas like the European VLBI Network (EVN) \citep{Charlot2004a}. This identification of new optically-bright VLBI sources for the Gaia frame alignment is the purpose of an observational program initiated in 2007 which focuses on 450 weak radio sources selected from the NVSS (NRAO VLA Sky Survey; \citealt{Condon1998}) and brighter than the apparent optical magnitude 18. In the initial step, about 90~\% of these weak radio sources have been detected during an EVN pilot experiment \citep{Bourda2008}, hence drawing good prospects for increasing the pool of VLBI extragalactic radio sources suitable for aligning the ICRF and the future Gaia frame with the highest accuracy.

\begin{figure}
\centering
\includegraphics[scale=0.35, angle=-90]{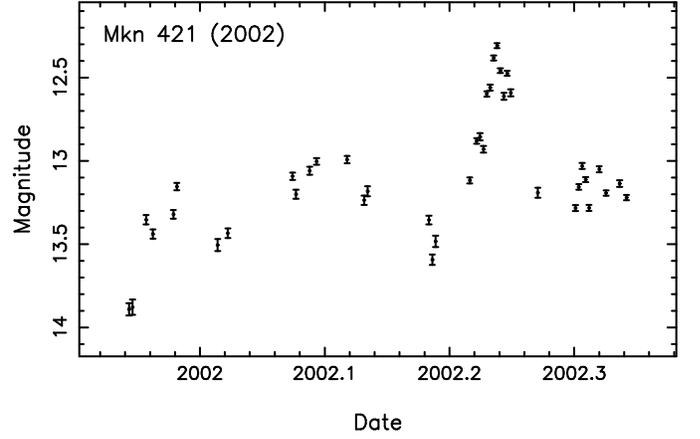}
\caption{Variation in the apparent optical magnitude of the BL Lac object Mkn\,421 ($1101+384$) from the ICRF, during the year 2002, from optical monitoring at Bordeaux Observatory \citep{Charlot2004b}.}
\label{Fig:Fig6}
\end{figure}

Another issue that might affect the choice of sources for this alignment is the possible variability of the apparent magnitude for the extragalactic sources observed at optical wavelengths. These variations are common and can easily reach one magnitude unit, especially for BL Lac objects that are known to vary on short time scales, as illustrated in Fig.~\ref{Fig:Fig6}. Such variability may affect the selection of candidate sources for the alignment of the ICRF and the Gaia frame if only a limited number of measurements are available for their optical magnitude. For this reason, it is desirable to engage in a long term optical monitoring on these candidate sources, especially those that have a magnitude near 18, to determine if on average they have an optical magnitude that is brighter or weaker than 18. While the results of this monitoring might impact at some level the list of sources in Table~\ref{table:tab3}, this should not significantly change the percentage of suitable ICRF sources currently available for the Gaia frame alignment.

Finally, it is important to note that obtaining the alignment between the ICRF and the future Gaia frame with the highest accuracy is essential, not only to ensure the consistency between the measured radio and optical positions, but also to pinpoint the relative location of the optical and radio emission in AGNs to a few tens of $\mu$as. In fact, the optical and radio emission may not always be spatially coincident at this level of accuracy due to opacity effects in AGN jets. For example, \citet{Kovalev2008} shows that on average the optical-radio core shift is at the level of 100~$\mu$as from studying a sample of 29 objects, which is significant considering the expected accuracy of the Gaia catalog and the one foreseen for the ICRF by 2015--2020. It will thus be important to use a large number of objects in aligning the two frames so that positional inconsistencies are averaged out. On the other hand, the differences between the optical and radio positions may provide a direct measurement of such core shifts, which would be of high interest for probing AGN jets properties.

%%%%%%%%%%%%%%%%%%%%%
\section{Conclusion}

This study focused on the astrometric suitability of the current ICRF extragalactic radio sources for the alignment with the future Gaia frame. We identified 243 candidate sources for this alignment, but only 70 of these (10\% of the ICRF) possess the high astrometric quality required to ensure the highest accuracy in the alignment. Accordingly, the current number of VLBI sources for accurately aligning the ICRF and the future Gaia frame is not sufficient. We also showed that the QSOs that will have the most accurate positions measured with Gaia are not those that have the best astrometric positions in the ICRF statistically.

To compensate for the lack of high astrometric quality extragalactic radio sources for this alignment, the VCS catalog, along with the future ICRF-2, will be of high interest. An additional direction would be to identify new appropriate VLBI sources from current deep radio surveys. These solutions mostly concern the northern hemisphere, since the VLBI arrays and observations are concentrated in this part of the world. However, to ensure a homogeneous sky coverage, a major effort is also necessary in the southern hemisphere, most specifically for the declinations below $-40^{\circ}$. To this end, the Asia Pacific Telescope \citep{Gulyaev2007} is likely to play a significant role.

%%%%%%%%%%%%%%%%%%%%%%%%%%
\begin{acknowledgements}
The first author is grateful to the CNES (Centre National d'Etudes Spatiales, France) for the post-doctoral position granted at Bordeaux Observatory. 
\end{acknowledgements}

%%%%%%%%%%%%%%%%%%%%%%%%%%

%%%%%%%%%%%%%%%%%%%%%%%%%%%%%%%%%%%%%%%%%%%%%%%%%%%%%%%%%%%%%%%%%%%%%%%%%%%%
\longtab{3}{
\begin{longtable}{llllrrrrrrlll}
\caption{\label{table:tab3} List of the 70 ICRF extragalactic radio sources suitable for the alignment with the future Gaia frame. The column labelled `SI' gives the X-band structure index of the source. The column labelled `Cat.' details if the source is categorized in the ICRF as \textit{defining} (d), \textit{candidate} (c), \textit{other} (o) or \textit{new} (n); and the columns labelled `VV2006 name', `Type' and `V' indicate respectively (i) the name of the source in the optical catalog V\'{e}ron \& V\'{e}ron (2006), (ii) the source type (quasar (Q), BL Lac object (B) or active galaxy (A)), and (iii) its apparent optical magnitude.} \\
\hline\hline
ICRF designation & IERS name & SI & Cat. & \multicolumn{3}{c}{$\alpha$} & \multicolumn{3}{c}{$\delta$} & VV2006 name & Type & V \\
   &   &   &   & h & min & sec & $^{\circ}$ & $^\prime$ & $^{\prime\prime}$ &   &   &   \\
\hline
\endfirsthead
\caption{continued.} \\
\hline\hline
ICRF designation & IERS name & SI & Cat. & \multicolumn{3}{c}{$\alpha$} & \multicolumn{3}{c}{$\delta$} & VV2006 name & Type & V \\
   &   &   &   & h & min & sec & $^{\circ}$ & $^\prime$ & $^{\prime\prime}$ &   &   &   \\
\hline
\endhead
\hline
\endfoot
J001031.0$+$105829 & 0007$+$106 & 1 & d &  0 & 10 & 31.0059 &     10 & 58 & 29.504 & III Zw  2        & A & 15.40 \\
J001331.1$+$405137 & 0010$+$405 & 2 & d &  0 & 13 & 31.1302 &     40 & 51 & 37.144 & 4C 40.01         & Q & 17.90 \\ 
J005041.3$-$092905 & 0048$-$097 & 1 & o &  0 & 50 & 41.3174 &   $-$9 & 29 &  5.210 & PKS 0048$-$09    & B & 17.44 \\
J011205.8$+$224438 & 0109$+$224 & 2 & d &  1 & 12 &  5.8247 &     22 & 44 & 38.786 & S2 0109$+$22     & B & 15.66 \\
J020504.9$+$321230 & 0202$+$319 & 2 & d &  2 &  5 &  4.9254 &     32 & 12 & 30.096 & DW 0202$+$31     & Q & 17.40 \\ 
J021046.2$-$510101 & 0208$-$512 & 2 & o &  2 & 10 & 46.2004 &  $-$51 &  1 &  1.892 & PKS 0208$-$512   & B & 16.93 \\
J021748.9$+$014449 & 0215$+$015 & 1 & d &  2 & 17 & 48.9547 &      1 & 44 & 49.699 & PKS 0215$+$015   & Q & 16.09 \\ 
J023838.9$+$163659 & 0235$+$164 & 1 & d &  2 & 38 & 38.9301 &     16 & 36 & 59.275 & AO 0235$+$164    & Q & 15.50 \\ 
J030335.2$+$471616 & 0300$+$470 & 2 & o &  3 &  3 & 35.2422 &     47 & 16 & 16.276 & 4C 47.08         & B & 16.95 \\
J031301.9$+$412001 & 0309$+$411 & 2 & d &  3 & 13 &  1.9621 &     41 & 20 &  1.184 & NRAO 128         & A & 18.00 \\
J033413.6$-$400825 & 0332$-$403 & 1 & o &  3 & 34 & 13.6545 &  $-$40 &  8 & 25.398 & PKS 0332$-$403   & B & 17.50 \\
J040534.0$-$130813 & 0403$-$132 & 1 & o &  4 &  5 & 34.0034 &  $-$13 &  8 & 13.691 & PKS 0403$-$13    & Q & 17.09 \\ 
J040748.4$-$121136 & 0405$-$123 & 2 & c &  4 &  7 & 48.4310 &  $-$12 & 11 & 36.660 & PKS 0405$-$12    & Q & 14.86 \\ 
J042446.8$+$003606 & 0422$+$004 & 2 & d &  4 & 24 & 46.8421 &      0 & 36 &  6.330 & PKS 0422$+$00    & B & 16.98 \\
J050842.3$+$843204 & 0454$+$844 & 2 & d &  5 &  8 & 42.3635 &     84 & 32 &  4.544 & S5 0454$+$84     & B & 16.50 \\
J045550.7$-$461558 & 0454$-$463 & 1 & c &  4 & 55 & 50.7725 &  $-$46 & 15 & 58.682 & PKS 0454$-$46    & Q & 16.90 \\ 
J050643.9$-$610940 & 0506$-$612 & 2 & d &  5 &  6 & 43.9887 &  $-$61 &  9 & 40.993 & PKS 0506$-$61    & Q & 16.85 \\ 
J053007.9$-$250329 & 0528$-$250 & 2 & o &  5 & 30 &  7.9628 &  $-$25 &  3 & 29.900 & PKS 0528$-$250   & Q & 17.34 \\ 
J060940.9$-$154240 & 0607$-$157 & 2 & c &  6 &  9 & 40.9495 &  $-$15 & 42 & 40.673 & PKS 0607$-$15    & Q & 18.00 \\ 
J064204.2$+$675835 & 0636$+$680 & 1 & d &  6 & 42 &  4.2574 &     67 & 58 & 35.621 & S4 0636$+$68     & Q & 16.60 \\ 
J072153.4$+$712036 & 0716$+$714 & 1 & d &  7 & 21 & 53.4485 &     71 & 20 & 36.363 & S5 0716$+$71     & B & 15.50 \\
J075706.6$+$095634 & 0754$+$100 & 2 & d &  7 & 57 &  6.6429 &      9 & 56 & 34.852 & PKS 0754$+$100   & B & 15.00 \\
J081126.7$+$014652 & 0808$+$019 & 1 & c &  8 & 11 & 26.7073 &      1 & 46 & 52.220 & PKS 0808$+$019   & B & 17.20 \\
J082601.5$-$223027 & 0823$-$223 & 1 & c &  8 & 26 &  1.5729 &  $-$22 & 30 & 27.204 & PKS 0823$-$223   & B & 16.20 \\
J083052.0$+$241059 & 0827$+$243 & 2 & c &  8 & 30 & 52.0862 &     24 & 10 & 59.820 & B2 0827$+$24     & Q & 17.26 \\ 
J083740.2$+$245423 & 0834$+$250 & 2 & n &  8 & 37 & 40.2457 &     24 & 54 & 23.122 & B2 0834$+$25     & Q & 17.90 \\ 
J091552.4$+$293324 & 0912$+$297 & 1 & d &  9 & 15 & 52.4016 &     29 & 33 & 24.043 & B2 0912$+$29     & B & 16.39 \\
J095533.1$+$690355 & 0951$+$693 & 1 & o &  9 & 55 & 33.1731 &     69 &  3 & 55.061 & NGC 3031         & Q & 11.63 \\ 
J095847.2$+$653354 & 0954$+$658 & 2 & d &  9 & 58 & 47.2451 &     65 & 33 & 54.818 & S4 0954$+$65     & B & 16.81 \\
J095820.9$+$322402 & 0955$+$326 & 2 & d &  9 & 58 & 20.9496 &     32 & 24 &  2.209 & 3C 232	          & Q & 15.78 \\
J101447.0$+$230116 & 1012$+$232 & 2 & d & 10 & 14 & 47.0654 &     23 &  1 & 16.571 & PKS 1011$+$23    & Q & 17.80 \\
J102444.8$+$191220 & 1022$+$194 & 2 & c & 10 & 24 & 44.8096 &     19 & 12 & 20.416 & 4C 19.34         & Q & 17.49 \\
J104423.0$+$805439 & 1039$+$811 & 2 & d & 10 & 44 & 23.0626 &     80 & 54 & 39.443 & S5 1039$+$81     & Q & 17.90 \\
J110427.3$+$381231 & 1101$+$384 & 2 & c & 11 &  4 & 27.3139 &     38 & 12 & 31.799 & MARK  421        & B & 12.90 \\
J110331.5$-$325116 & 1101$-$325 & 2 & c & 11 &  3 & 31.5264 &  $-$32 & 51 & 16.692 & PKS 1101$-$325   & Q & 16.30 \\
J111358.6$+$144226 & 1111$+$149 & 2 & d & 11 & 13 & 58.6951 &     14 & 42 & 26.953 & PKS 1111$+$149   & Q & 17.90 \\
J112553.7$+$261019 & 1123$+$264 & 2 & o & 11 & 25 & 53.7119 &     26 & 10 & 19.979 & PKS 1123$+$26    & Q & 18.00 \\
J114658.2$+$395834 & 1144$+$402 & 1 & o & 11 & 46 & 58.2979 &     39 & 58 & 34.305 & S4 1144$+$40     & Q & 18.00 \\
J115019.2$+$241753 & 1147$+$245 & 2 & d & 11 & 50 & 19.2122 &     24 & 17 & 53.835 & B2 1147$+$24     & B & 15.74 \\
J121752.0$+$300700 & 1215$+$303 & 2 & d & 12 & 17 & 52.0820 &     30 &  7 &  0.636 & B2 1215$+$30     & B & 15.62 \\
J122222.5$+$041315 & 1219$+$044 & 2 & d & 12 & 22 & 22.5496 &      4 & 13 & 15.776 & PKS 1219$+$04    & Q & 17.98 \\
J122503.7$+$125313 & 1222$+$131 & 2 & n & 12 & 25 &  3.7433 &     12 & 53 & 13.139 & NGC 4374         & A & 12.31 \\
J125438.2$+$114105 & 1252$+$119 & 2 & d & 12 & 54 & 38.2556 &     11 & 41 &  5.895 & PKS 1252$+$11    & Q & 16.64 \\
J133245.2$+$472222 & 1330$+$476 & 1 & n & 13 & 32 & 45.2464 &     47 & 22 & 22.668 & B3 1330$+$476    & Q & 17.96 \\
J135256.5$-$441240 & 1349$-$439 & 2 & c & 13 & 52 & 56.5349 &  $-$44 & 12 & 40.387 & PKS 1349$-$439   & B & 16.37 \\
J141908.1$+$062834 & 1416$+$067 & 2 & d & 14 & 19 &  8.1802 &      6 & 28 & 34.803 & 3C 298.0         & Q & 16.79 \\
J142230.3$+$322310 & 1420$+$326 & 1 & c & 14 & 22 & 30.3790 &     32 & 23 & 10.440 & B2 1420$+$32     & Q & 17.50 \\
J142700.3$+$234800 & 1424$+$240 & 2 & c & 14 & 27 &  0.3918 &     23 & 48 &  0.038 & PKS 1424$+$240   & B & 14.95 \\
J151053.5$-$054307 & 1508$-$055 & 2 & n & 15 & 10 & 53.5914 &   $-$5 & 43 &  7.417 & PKS 1508$-$05    & Q & 17.21 \\
J152237.6$-$273010 & 1519$-$273 & 1 & c & 15 & 22 & 37.6760 &  $-$27 & 30 & 10.785 & PKS 1519$-$273   & B & 17.70 \\
J154049.4$+$144745 & 1538$+$149 & 2 & d & 15 & 40 & 49.4915 &     14 & 47 & 45.885 & 4C 14.60         & Q & 17.30 \\
J154929.4$+$023701 & 1546$+$027 & 2 & c & 15 & 49 & 29.4368 &      2 & 37 &  1.164 & PKS 1546$+$027   & Q & 17.45 \\
J163813.4$+$572023 & 1637$+$574 & 2 & d & 16 & 38 & 13.4563 &     57 & 20 & 23.979 & OS 562	          & Q & 16.90 \\
J172824.9$+$042704 & 1725$+$044 & 2 & d & 17 & 28 & 24.9527 &      4 & 27 &  4.914 & PKS 1725$+$044   & Q & 16.99 \\
J172818.6$+$501310 & 1727$+$502 & 2 & d & 17 & 28 & 18.6239 &     50 & 13 & 10.470 & I Zw 187         & B & 15.97 \\
J175132.8$+$093900 & 1749$+$096 & 1 & c & 17 & 51 & 32.8186 &      9 & 39 &  0.729 & OT 081	          & Q & 16.78 \\
J180132.3$+$440421 & 1800$+$440 & 1 & d & 18 &  1 & 32.3149 &     44 &  4 & 21.900 & B3 1800$+$440    & Q & 17.90 \\
J184916.0$+$670541 & 1849$+$670 & 1 & d & 18 & 49 & 16.0723 &     67 &  5 & 41.680 & S4 1849$+$67     & Q & 16.90 \\
J192332.1$-$210433 & 1920$-$211 & 2 & c & 19 & 23 & 32.1898 &  $-$21 &  4 & 33.333 & TEX 1920$-$211   & Q & 17.50 \\
J195759.8$-$384506 & 1954$-$388 & 2 & d & 19 & 57 & 59.8193 &  $-$38 & 45 &  6.356 & PKS 1954$-$388   & Q & 17.07 \\
J210138.8$+$034131 & 2059$+$034 & 2 & d & 21 &  1 & 38.8342 &      3 & 41 & 31.322 & PKS 2059$+$034   & Q & 17.78 \\ 
J212912.1$-$153841 & 2126$-$158 & 2 & c & 21 & 29 & 12.1759 &  $-$15 & 38 & 41.041 & PKS 2126$-$15    & Q & 17.00 \\ 
J214622.9$-$152543 & 2143$-$156 & 2 & d & 21 & 46 & 22.9793 &  $-$15 & 25 & 43.885 & PKS 2143$-$156   & Q & 17.27 \\ 
J215852.0$-$301332 & 2155$-$304 & 2 & c & 21 & 58 & 52.0651 &  $-$30 & 13 & 32.118 & PKS 2155$-$304   & B & 13.09 \\
J222940.0$-$083254 & 2227$-$088 & 1 & c & 22 & 29 & 40.0843 &   $-$8 & 32 & 54.435 & PKS 2227$-$08    & Q & 17.43 \\ 
J225307.3$+$194234 & 2250$+$190 & 2 & n & 22 & 53 &  7.3692 &     19 & 42 & 34.629 & HS 2250$+$1926   & Q & 18.00 \\ 
J225717.5$+$024317 & 2254$+$024 & 1 & c & 22 & 57 & 17.5631 &      2 & 43 & 17.512 & PKS 2254$+$024   & Q & 17.96 \\ 
J225717.3$+$074312 & 2254$+$074 & 1 & d & 22 & 57 & 17.3031 &      7 & 43 & 12.303 & PKS 2254$+$074   & B & 16.36 \\
J225805.9$-$275821 & 2255$-$282 & 2 & o & 22 & 58 &  5.9629 &  $-$27 & 58 & 21.257 & PKS 2255$-$282   & Q & 16.77 \\ 
J230343.5$-$680737 & 2300$-$683 & 2 & c & 23 &  3 & 43.5663 &  $-$68 &  7 & 37.462 & PKS 2300$-$683   & Q & 16.38 \\ 
\end{longtable}
}

\end{document}